\documentclass[prb, preprint]{revtex4-1}
\usepackage{graphicx}
\usepackage{amsmath}
\usepackage{amsfonts}
\usepackage{siunitx} 
\usepackage{bm}

\begin{document}
\title{An ideal Fermi gas under uniform gravity}
\author{Pattarapon Tanalikhit}
\affiliation{Department of Physics, Korea Advanced Institute of Science and Technology, Daejeon 34141, Korea}
\author{Wittaya Kanchanapusakit}
\affiliation{Department of Physics, King Mongkut's University of Technology Thonburi, Pracha Uthit Road, Bangkok 10140, Thailand}

% \date{\today}
	
\begin{abstract}
We consider an ideal Fermi gas in a container subject to a uniform gravitational field at absolute zero temperature. Under a semiclassical approximation, we examine the density profile of the particles and derive an expression for the chemical potential. A critical value of the chemical potential separates the weak- and strong-gravity regimes, and the kinetic and potential energies of the Fermi gas are determined in both regimes.

\end{abstract}

\vspace{2pc}

\maketitle

\section{Introduction}

The Fermi gas at absolute zero temperature is a fundamental model in quantum mechanics and statistical physics, describing a system of noninteracting fermions where all quantum states are filled up to the Fermi energy due to the Pauli exclusion principle. It provides an essential understanding of the behavior of electrons in metals, explaining properties such as electrical conductivity and specific heat. The Fermi gas also forms the basis of the concept of degeneracy pressure, which plays a crucial role in supporting compact astrophysical objects such as white dwarfs and neutron stars.\cite{Silbar_2004, Potekhin_2010} Its simplicity and accuracy make the Fermi gas model essential for studying a wide range of quantum systems at their ground state.

A previous study analyzes the Fermi gas under Earth’s gravity in a semi-infinite space analogous to an atmosphere, i.e., for $0\leq z< \infty$, where $z$ denotes the distance measured from the ground.\cite{Kaneko} In contrast, this article explores the effect of an upper boundary by considering the gas confined within a container.\cite{footnote1} Confinement in the container modifies the thermodynamic properties of the system, particularly their dependence on gravity. The analysis proceeds in sections as follows. Section II briefly reviews the fundamental properties of the Fermi gas without gravity. Section III considers a uniform gravitational field and derives the resulting density profiles within the semiclassical framework. In Sec. IV, the chemical potential is examined in the weak- and strong-gravity regimes, with their boundary defining the critical gravity. The results are used to calculate the kinetic and potential energies of the system. The discussion in Sec. V includes the exact quantum problem, some numerical results, and a verification of the energy relations using the virial theorem.

\section{Ideal Fermi gas at $T=0$}

\begin{figure}[h]	
\centering
\includegraphics[width=8.5cm]{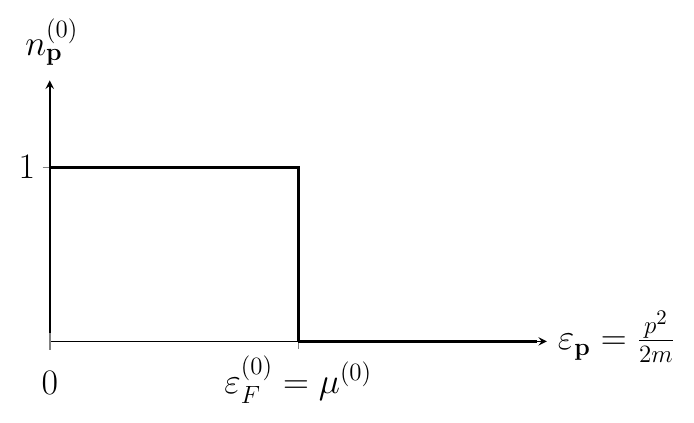}
\caption{At $T=0$, the Fermi-Dirac distribution function is a step function, where all states with energy lower than $\varepsilon_F^{(0)}$ are occupied and all states with energy higher than $\varepsilon_F^{(0)}$ are empty.}
\label{dist_function_free}
\end{figure}

Consider $N$ noninteracting Fermi particles, each with mass $m$ and momentum $\mathbf{p}$, in a container of a fixed volume $V$ at temperature $T$. The probability that a particle occupies a certain energy level at temperature $T$ is given by the Fermi-Dirac distribution function. At $T=0$, the distribution function, denoted by $n_\mathbf{p}^{(0)}$, as a function of the particle energy $\varepsilon_{\mathbf{p}}$  is shown in Fig. \ref{dist_function_free} (we use the superscript $(0)$ to refer to the quantity in the absence of an external field). All energy levels up to the Fermi energy, $\varepsilon_F^{(0)}$, are fully occupied, while levels above $\varepsilon_F^{(0)}$ are unoccupied. The Fermi energy is equal to the chemical potential $\mu^{(0)}$. By summing over $\mathbf{p}$, the number density of the particles can be determined:
\begin{equation}
    \frac{N}{V} = \sum_{\mathbf{p}}n_{\mathbf{p}}^{(0)} = 2\int_0^{\sqrt{2m\varepsilon_F^{(0)}}} \frac{d^3 p}{(2\pi\hbar)^3} = \frac{1}{3\pi^2\hbar^3}\left(2m\varepsilon_F^{(0)}\right)^{3/2},
\label{density_0}
\end{equation}
where the summation is replaced by $2\int d^3p/(2\pi\hbar)^3 = 2\int 4\pi p^2dp/(2\pi\hbar)^3$, and the spin degeneracy factor of 2 has been included. Given $N$ and $V$, Eq. (\ref{density_0}) provides the expressions for the Fermi energy and the chemical potential:
\begin{equation}
    \varepsilon_F^{(0)} = \mu^{(0)} = \frac{\hbar^2}{2m}\left(3\pi^2 \frac{N}{V} \right)^{2/3}.
\label{E_F_0}
\end{equation}

Knowing the distribution function $n_\mathbf{p}^{(0)}$, one can find the kinetic energy density (kinetic energy per unit volume) 
\begin{equation}
    K^{(0)}_V = \sum_{\mathbf{p}}n_{\mathbf{p}}^{(0)}\frac{p^2}{2m} = 2\int_0^{\sqrt{2m\varepsilon_F^{(0)}}} \frac{d^3 p}{(2\pi\hbar)^3} \frac{p^2}{2m} = \frac{1}{10\pi^2m\hbar^3}\left(2m\varepsilon_F^{(0)}\right)^{5/2}.
\label{energy_0}
\end{equation}
The total kinetic energy is $K^{(0)}=K_{V}^{(0)}V$. It follows from Eqs. (\ref{density_0}) and (\ref{energy_0}) that $K^{(0)}=3N\varepsilon_F^{(0)}/5$.

\section{Density profile under gravity}

With a uniform downward gravitational field of magnitude $g$, the number density $n(z)$ becomes dependent on the height $z$, measured from the base of the container. The particle number $N$ and the container volume $V$ are held constant, while $g$ is treated as a variable parameter. This section investigates the shape of the density profile and how it changes with $g$.

A good approximation of the ground-state density distribution of fermions in an external potential can be obtained using the Thomas-Fermi approximation,\cite{Pethick} which has previously been applied to a harmonically trapped Fermi gas.\cite{Butts}  Within this semiclassical framework, the gas at position $\mathbf{r}$ is treated as locally uniform, with its properties determined by a uniform system whose density equals the local density $n(\mathbf{r})$. Under the uniform gravity, it follows from Eq. (\ref{E_F_0}) that the local Fermi energy, as a function of $z$, is related to the local density:
\begin{equation}
    \varepsilon_F(z) = \frac{\hbar^2}{2m}\left[3\pi^2 n(z) \right]^{2/3}.
\label{E_F_local}
\end{equation}
By definition, the chemical potential $\mu$ is the energy required to add a particle at a given point in the container. In the presence of an external gravitational field, this energy is given by the sum of the local Fermi energy $\varepsilon_F(z)$ and the potential energy $mgz$, $\mu = \varepsilon_F(z)+mgz$, where the zero of the potential is chosen at the base of the container. The condition for diffusive equilibrium is that $\mu$ remains constant for any $z$.\cite{Stat_Phys_part_1_mu} This implies that an increase in gravitational potential energy with $z$ must be offset by a reduction in $\varepsilon_F(z)$, corresponding to a decrease in $n(z)$. 

\begin{figure}[t]	
\centering
\includegraphics[width=8.5cm]{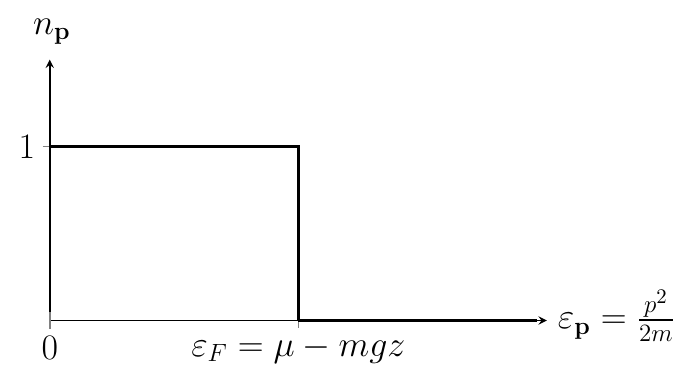}
\caption{In the presence of the gravitational field, all states with energy lower than $\varepsilon_F=\mu-mgz$ are occupied, and all states above $\varepsilon_F$ are empty.}
\label{dist_function_gravity}
\end{figure}

\begin{figure}[t]	
\centering
\includegraphics[width=8.5cm]{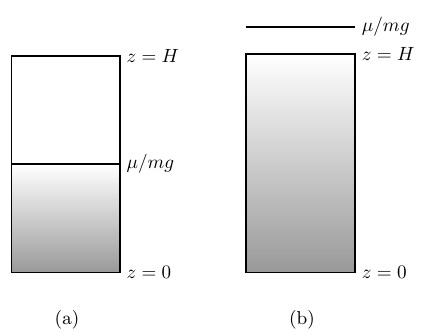}
\caption{For a container with height $H$, the coloured area represents the space occupied by the particles in cases where (a) $\mu/mg < H$ and (b) $\mu/mg > H$. The color gradient represents the density profile, with darker shading indicating higher density.}
\label{particles}
\end{figure}

The highest occupied energy level at a given $z$ is given by
\begin{equation}
    \varepsilon_F(z)=\mu-mgz.
\label{E_F}
\end{equation}
As shown in Fig. \ref{dist_function_gravity}, the Fermi-Dirac distribution function remains a step function, but the discontinuity takes place at $\varepsilon_{\mathbf{p}}=\varepsilon_F(z)$. At height $z$, the number density can be calculated as in Eq. (\ref{density_0}) but with the upper limit of $\sqrt{2m\varepsilon_F(z)}=\sqrt{2m(\mu-mgz)}$. Equivalently, according to Eqs. (\ref{E_F_local}) and (\ref{E_F}), the density profile is 
\begin{eqnarray}
    n(z) = 
    \begin{cases}\dfrac{1}{3\pi^2\hbar^3}\left[2m(\mu-mgz) \right]^{3/2} &z < \mu/mg, \\
    0 &z>\mu/mg.
    \end{cases}
\label{density_gravity}
\end{eqnarray}
This equation indicates that the density decreases with height and becomes zero if $z \ge \mu/mg$. Consider a cylindrical container with cross-section $A$ and height $H$, as depicted in Fig. \ref{particles}. If $\mu/mg<H$, the particles occupy only a fraction of the container's volume. However, if $\mu/mg>H$, the density decreases with height but never reaches zero. 

Equation (\ref{density_gravity}) can be expressed in terms of $N$ via the constraint $\int_{0}^{z'}n(z)Adz=N$, where $z'$ is the maximum height reached by the particles. As illustrated in Fig. \ref{particles}, $z'=
\mu/mg$ if $\mu/mg < H$, and $z'=H$ if $\mu/mg>H$. Inserting Eq. (\ref{density_gravity}) into the integral, evaluating, and re-arranging, yields:
\begin{equation}
\frac{n(z)}{N/V} = \dfrac{5}{2}\dfrac{mgH}{\mu}\left(1-\dfrac{mgz}{\mu}\right)^{3/2}
\times\begin{cases}1 &\mu/mg<H,\\  \dfrac{1}{1-\left(1-mgH/\mu\right)^{5/2}} &\mu/mg>H.\end{cases}
\label{n_z}
\end{equation}
The upper case is relevant for $0\leq z\leq \mu/mg$ in Fig. \ref{particles}a. The lower case is valid for $0\leq z\leq H$ in Fig. \ref{particles}b.

\section{Dependence of energies on gravity}

\subsection{Chemical potential}

By letting $z=0$ in Eq. (\ref{E_F}) and using Eq. (\ref{E_F_local}), one arrives at $\mu=\varepsilon_F(0)=\hbar^2\left[3\pi^2n(0)\right]^{2/3}/2m$. Relative to the chemical potential in the absence of the field, $\mu^{(0)}$ as given by Eq. (\ref{E_F_0}), $\mu$ in the presence of the field can be expressed as
\begin{equation}
    \frac{\mu}{\mu^{(0)}} = \left(\frac{n(0)}{N/V} \right)^{2/3}.
\label{mu_gravity}
\end{equation}
Inserting Eq. (\ref{n_z}) (evaluated at $z = 0$) and re-arranging, we find:
\begin{equation}
    \frac{\mu}{\mu^{(0)}}=\begin{cases}\left(\dfrac{5}{2}\dfrac{mgH}{\mu^{(0)}}\right)^{2/5} &\mu/mg < H, \\
    \left[\dfrac{5mgH/2\mu^{(0)}}{1-(1-mgH/\mu)^{5/2}}\right]^{2/5} &\mu/mg>H.\end{cases}
\label{mu_gravity}
\end{equation}
Since $\mu^{(0)}$ is a constant independent of the gravitational field, Eq. (\ref{mu_gravity}) determines how $\mu$ varies with $g$. The upper case, $\mu \propto g^{2/5}$, is consistent with the result of Kaneko et al.,\cite{Kaneko} corresponding to the strong-gravity regime in which the particles occupy only part of the container (Fig. \ref{particles}a). In this article, we extend that result by reporting the opposite regime, the lower case of Eq. (\ref{mu_gravity}), where gravity is sufficiently weak that particle density is non-zero throughout the container (Fig. \ref{particles}b). 

\begin{figure}[t]	
\centering
\includegraphics[width=8.5cm]{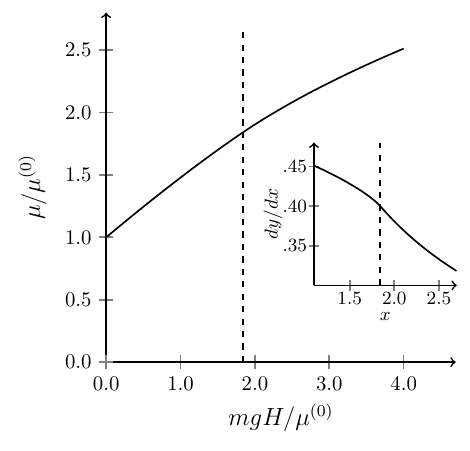}
\caption{Dependence of $\mu$ on $g$ for a fixed $H$. The dashed line indicates the critical point $mgH/\mu^{(0)}=1.842$, where the two cases in Eq. (\ref{mu_gravity}) meet. The inset plot shows the first derivative of $y=\mu/\mu^{(0)}$ with respect to $x=mgH/\mu^{(0)}$, indicating the differentiability at the critical point.}
\label{mu_vs_gravity}
\end{figure}

We numerically solve the lower case of Eq. (\ref{mu_gravity}) in Fig. \ref{mu_vs_gravity}; the two regimes connect smoothly at the critical point $mgH/\mu^{(0)} = (5/2)^{2/3}= 1.842$. This critical point defines the critical gravity. For a fixed $H$, the strong-gravity regime is defined by $mgH/\mu^{(0)} > 1.842$, while the weak-gravity regime corresponds to $mgH/\mu^{(0)} < 1.842$. It is observed that the initial slope in Fig. \ref{mu_vs_gravity} is $1/2$. This behavior can be confirmed by applying a binomial expansion to Eq.~(\ref{mu_gravity}) in the low-gravity limit $mgH/\mu^{(0)} \ll 1$, which gives
$\mu/\mu^{(0)} \approx 1 + mgH/2\mu^{(0)}$.

\subsection{Kinetic and potential energy}

The Fermi-Dirac distribution function, illustrated in Fig. \ref{dist_function_gravity}, provides the framework for calculating the kinetic energy density:
\begin{align}
    K_V &= \sum_{\mathbf{p}}n_{\mathbf{p}}\frac{p^2}{2m} = 2\int_0^{\sqrt{2m(\mu-mgz)}} \frac{d^3 p}{(2\pi\hbar)^3} \frac{p^2}{2m} \nonumber \\
    &=\frac{1}{10\pi^2\hbar^3m}\left[2m(\mu-mgz)\right]^{5/2}.
\label{K_density_gravity}
\end{align}
Given the density $n(z)$, the potential energy density at a height $z$ is given by
\begin{equation}
    U_V = mgzn(z).
\label{U_density_gravity}
\end{equation}
By integrating Eqs. (\ref{K_density_gravity}) and (\ref{U_density_gravity}) over the volume occupied by the particles, the total kinetic energy $K=\int_0^{z'}K_V Adz$ and potential energy $U=\int_0^{z'} U_V Adz$ can be determined. However, it is important to consider the two scenarios in Fig. \ref{particles}: when $\mu/mg < H$, the particles reach a maximum height of $z' = \mu/mg$, whereas for $\mu/mg > H$, the maximum height is $z' = H$.

\begin{figure}[t]	
\centering
\includegraphics[width=8.5cm]{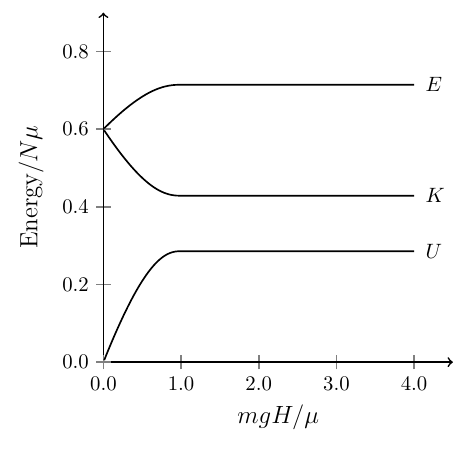}
\caption{The energies in the unit of $N\mu$ as the parameter $mgH/\mu$ varies. If $mgH/\mu \geq 1$, $K=3N\mu/7$, $U=2N\mu/7$ and $E=5N\mu/7$.}
\label{Energy_Nmu}
\end{figure}

We can express the energies in terms of the number of particles $N=\int_0^{z'} n(z)Adz$. From Eqs. (\ref{density_gravity}) and  (\ref{K_density_gravity}), the kinetic energy per particle can be written as
\begin{equation}
    \frac{K}{N} = \frac{3}{5}\frac{\int_0^{z'}(\mu-mgz)^{5/2}dz}{\int_0^{z'}(\mu-mgz)^{3/2}dz}.
\end{equation}
Choosing an appropriate upper limit $z'$, we obtain 
\begin{equation}
    K = \begin{cases}\dfrac{3}{7}N\mu &\mu/mg<H, \\ 
    \dfrac{3}{7}N\mu\left[\dfrac{1-(1-mgH/\mu)^{7/2}}{1-(1-mgH/\mu)^{5/2}}\right] &\mu/mg>H. \end{cases}
\label{K_gravity}
\end{equation}
Similarly, using Eqs. (\ref{density_gravity}) and (\ref{U_density_gravity}), the potential energy per particle becomes
\begin{equation}
    \frac{U}{N}=mg\,\frac{\int_0^{z'}z(\mu-mgz)^{3/2}dz}{\int_0^{z'}(\mu-mgz)^{3/2}dz}.
\end{equation}
With a suitable $z'$, the potential energy is given by 
\begin{equation}
 U = \begin{cases}\dfrac{2}{7}N\mu &\mu/mg<H, \\
    \dfrac{2}{7}N\mu \left[\dfrac{1-(1-mgH/\mu)^{7/2}}{1-(1-mgH/\mu)^{5/2}}\right] 
    -NmgH\left[\dfrac{(1-mgH/\mu)^{5/2}}{1-(1-mgH/\mu)^{5/2}}\right] &\mu/mg>H. 
    \end{cases}
\label{U_gravity}
\end{equation}
Once $K$ and $U$ have been established, the total energy is $E=K+U$. In the unit of $N\mu$, these energies are presented graphically in Fig. \ref{Energy_Nmu} as functions of $mgH/\mu$. 

\begin{figure}[t]	
\centering
\includegraphics[width=8.5cm]{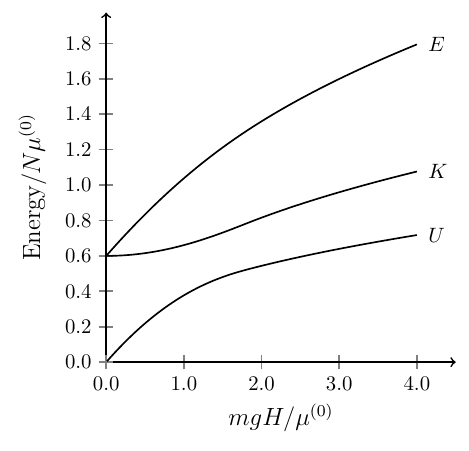}
\caption{The gravity dependence of the energies. The energies are measured relative to $N\mu^{(0)}$.}
\label{Energy_vs_gravity}
\end{figure}

Figure \ref{Energy_Nmu} does not exactly represent how the energies vary with $g$ because $\mu$ also changes with $g$. To fully account for the effect of $g$ on the energies, we must consider the relationship between $\mu$ and $g$, as illustrated in Fig. \ref{mu_vs_gravity}. By incorporating this relationship into the expressions for $K$, $U$ and $E$, the variation of these energies with $g$ can be derived, as shown in Fig. \ref{Energy_vs_gravity}. For given $N$ and $V$, the greater the gravity, the higher the energies. In the case of extremely weak gravity, where $mgH/\mu^{(0)} \ll 1$, we have $K\approx 3N\mu^{(0)}/5$ and $U\approx NmgH/2$, corresponding to a scenario where the particles are nearly uniformly distributed in the container, with the center of mass close to $z=H/2$.

\section{Discussion}

\subsection{Exact quantum problem and relation to the semiclassical description}

We consider noninteracting fermions in a uniform gravitational field $V(z)=mgz$ confined in a container of height $H$. The motion of a single particle along $z$ obeys the one--dimensional Schr\"odinger equation
\begin{equation}
    -\frac{\hbar^2}{2m}\frac{d^2\psi}{dz^2}+mgz\,\psi=\varepsilon\,\psi,
\label{Scrodinger}
\end{equation}
with the boundary conditions $\psi(0)=\psi(H)=0$.
This equation can be put into the Airy form:
\begin{equation}
    \frac{d^2\psi}{d\xi^2}=\xi\,\psi,
\label{Airy}
\end{equation}
where $\xi=(z-z')/\ell_g$, with 
\begin{equation}
z'=\frac{\varepsilon}{mg}, \qquad \ell_g=\left(\frac{\hbar^2}{2m^2g}\right)^{1/3}.
\end{equation}
The parameter $z'$ is the classical turning point defined by the vanishing of the local kinetic energy $\varepsilon-mgz$. The parameter $\ell_g$ is the characteristic gravitational length. 

The solution to Eq. (\ref{Airy}) is in the form of Airy functions. In the classically allowed region ($z<z'$), $\psi$ oscillates with increasing wavelength, while in the classically forbidden region ($z>z'$) it decays exponentially.\cite{Langhoff, Banacloche} Because the system is translationally invariant in the horizontal plane, the overall wave function factorizes as $\Psi_{n,\mathbf k_\perp}(\mathbf r)=e^{i\mathbf k_\perp\cdot\mathbf r_\perp}\psi_n(z)$ and the energy spectrum separates into
\begin{equation}
    \varepsilon_{n,\mathbf k_\perp} = \varepsilon_n + \frac{\hbar^2 k_\perp^2}{2m},
\end{equation}
where $\varepsilon_n$ arises from Airy quantization along $z$, and the second term represents the kinetic energy associated with free motion in the horizontal plane.

To pass from the single-particle spectrum to the many-body problem, we construct the grand canonical ensemble for an ideal Fermi gas, in which each single-particle state $(n,\mathbf k_\perp)$ is occupied according to the Fermi-Dirac distribution $f(\varepsilon_{n,\mathbf k_\perp}-\mu)$, with $f(x)=1/(e^{\beta x}+1)$ and $\beta=1/k_B T$. The exact grand potential is
\begin{equation}
    \Omega=-k_BT\sum_{n,\mathbf k_\perp}
    \ln{\left[1+e^{-\beta(\varepsilon_{n,\mathbf k_\perp}-\mu)}\right]},
\end{equation}
and the particle density is
\begin{equation}
    n(z)=\sum_{n,\mathbf k_\perp}
    f(\varepsilon_{n,\mathbf k_\perp}-\mu)\,|\psi_n(z)|^2.
\end{equation}
The classical turning point relevant for the many-body density corresponds to the highest occupied energy, $\varepsilon=\mu$ (at $T=0$), which gives $z'=\mu/mg$. The exact quantum treatment reveals that the wave function, and hence the particle number density, remains finite at the turning point $z=z'$ and extends into the classically forbidden region. This contrasts with the semiclassical result in Eq. (\ref{density_gravity}), where the density vanishes for $z\geq z'$, as depicted in Fig. \ref{particles}.

The conditions for the semiclassical approximation are (i) the local density $n(z)$ must be large enough to ensure that the sums over levels can be replaced by integrals over phase space, and (ii) the external potential must vary slowly on the scale of the local Fermi wavelength. Condition (i) is likely to break down near the classical turning point $z=z'$ where the density may be small. In the limiting case of weak gravity, $z'\gg H$ (Fig. \ref{particles}b), the classical turning point is above the ceiling boundary. Therefore, the local density remains sufficiently large for the semiclassical approximation to be valid throughout the system. In the opposite case of strong gravity, $z'<H$ (Fig. \ref{particles}a), and the turning point lies within the container. The local density becomes small in the vicinity of $z=z'$, and the semiclassical approximation breaks down here. In this regime, quantum effects near the classical turning point become significant and are governed by the Airy-function structure of the exact solution. In a narrow region below the turning point, the density exhibits oscillatory behavior in this classically allowed region, which, after summation over occupied states, leads to Friedel-type spatial oscillations.\cite{Friedel1952} In a narrow region above the turning point, the density decays exponentially and forms a finite quantum tail in the classically forbidden region. The semiclassical approximation treats this region as strictly forbidden and predicts a sharp cutoff of the density, and therefore fails to describe the correct behavior on both sides of the turning point.

\subsection{Critical gravity}

For given $N$, $V$, and $H$, Fig. \ref{mu_vs_gravity} shows the boundary between the weak- and the strong-gravity regimes, defined by the critical gravity
\begin{equation}
    g_{\textrm{crit}}=1.842 \frac{\mu^{(0)}}{mH}.
\end{equation}
The critical gravity is not an intrinsic property of the gas alone, but depends explicitly on the container height. Table \ref{table_1} shows a comparison of $H$, $N/V$, $\mu^{(0)}$, and $g_{\textrm{crit}}$ of degenerate non-relativistic fermionic matter in metals, white dwarfs, and neutron stars. The parameter $H$ is chosen to represent a macroscopic length scale, corresponding to a typical length for metals and to the characteristic radius of the stars.\cite{Koester} The number density is estimated by dividing the mass density by the atomic mass. The chemical potential $\mu^{(0)}$ is determined by using Eq. (\ref{E_F_0}). It can be seen that the critical gravity is of the order $ 10^{10}-10^{12}$ m.s$^{-2}$ for these types of matter. 

When compared with Earth’s gravitational field strength, metals on Earth lie in the low-gravity regime; the effect of gravity on the particle distribution is negligible. A typical white dwarf star with a mass of one solar mass has a surface gravity of the order $10^{6}$ m.s$^{-2}$, placing it also in the weak-gravity regime. Neutron stars possess a much stronger gravitational field, with $g \sim 10^{12}$ m.s$^{-2}$ on the surface, and thus lie in the vicinity of the critical field, where gravity begins to significantly influence the particle distribution.

\begin{table}[h!]
\centering
\caption{Order-of-magnitude estimates}
\begin{ruledtabular}
\begin{tabular}{l c c c c c}
% The codes above determine the horizontal alignment in each column.
% Options are l (left), r (right), c (centered), and p (paragraph).
% The p option allows an entry to be broken into multiple lines, and
% therefore requires a width specification, in this case 5 centimeters.
Matter & Particle & $H$ (m) & $N/V$ (m$^{-3}$) & $\mu^{(0)}$ (eV) &$g_{\textrm{crit}}$ (m.s$^{-2}$) \\
\hline	% horizontal line to separate headings from data
Metal & Electron & \ \ 1 & $10^{29}$ & 10 & $10^{12}$ \\
White dwarf & Electron & \ \ $10^6$ & $10^{36}$ & $10^5$ & $10^{10}$ \\
Neutron star & Neutron & \ \ $10^4$ & $10^{43}$ & $10^{7}$ & $10^{11}$ \\
\end{tabular}
\end{ruledtabular}
\label{table_1}
\end{table}

\subsection{Virial theorem}

After obtaining the results for $K$ in Eq. (\ref{K_gravity}) and $U$ in Eq. (\ref{U_gravity}), we want to verify these results against the virial theorem. Firstly, we need to find the pressure profile. As with any fluid in a gravitational field, the pressure must change with height to maintain hydrostatic equilibrium. The pressure function $P(z)$ satisfies \cite{Bradley}
\begin{equation}
    \frac{d}{dz}P(z) = -mgn(z).
\label{hydrostatic}
\end{equation}
To solve for $P(z)$ by integration, Eq. (\ref{n_z}) is used with suitable ranges of $z$ as shown in Fig. \ref{particles}. For $\mu/mg<H$, the range of $z$ is $0\leq z \leq \mu/mg$ with a boundary condition $P(\mu/mg)=0$. For $\mu/mg>H$, the range of $z$ is $0\leq z\leq H$; the pressure is extrapolated to meet the condition $P(\mu/mg)=0$. Therefore,
\begin{align}
    P(z) = &\frac{Nmg}{A}\left(1-\frac{mgz}{\mu}\right)^{5/2}  \times \begin{cases}1 &\mu/mg <H,\\
    \dfrac{1}{1-(1-mgH/\mu)^{5/2}} &\mu/mg>H. \end{cases}
\label{P_z}
\end{align}
where $A=V/H$ is the cross-sectional area of the container. 

Next, consider a system of particles confined within a surface $S$. Given that the potential energy is a homogeneous function of degree $\alpha$ in the coordinates, the virial theorem states that $2K-\alpha U = \int_S P(\mathbf{r})\mathbf{r}\cdot\hat{\mathbf{n}}dS$ where $\mathbf{r}$ is the position vector, and $\hat{\mathbf{n}}$ is the unit vector of the surface element $dS$ that experiences the pressure $P(\mathbf{r})$.\cite{Stat_Phys_part_1, Goldstein} For particles under uniform gravity, it implies that $\alpha=1$, and the virial theorem reads
\begin{equation}
	2K-U  =  \int_S P(\mathbf{r}) \mathbf{r}\cdot\hat{\mathbf{n}}dS.
\label{virial}
\end{equation}
A prior study on an ideal gas in a cylindrical container under uniform gravity shows that the above surface integral can be expressed as \cite{Kanchanapusakit}
\begin{equation}
    \int_S P(\mathbf{r}) \mathbf{r}\cdot\hat{\mathbf{n}}dS= 2\int_0^{z'}P(z)dV + P(z')V,
\label{surface_integral}
\end{equation}
where $dV=Adz$, and $z'$, the upper integration limit, is the maximum height reached by the particles. 

The previously obtained $K$, $U$, and $P(z)$ provide an opportunity to test the virial theorem. For $\mu/mg <H$, the upper integration limit is $z'=\mu/mg$. It follows from Eqs. (\ref{P_z}) and (\ref{surface_integral}) that
\begin{align}
    \int_S P(\mathbf{r}) \mathbf{r}\cdot\hat{\mathbf{n}}dS &= 2\int_0^{\mu/mg}P(z)dV + P(\mu/mg)V \nonumber \\
    &=2A\int_0^{\mu/mg}P(z)dz \nonumber \\
    &=2A\int_0^{\mu/mg}\frac{Nmg}{A}\left(1-\frac{mgz}{\mu}\right)^{5/2}dz \nonumber \\
    &=\frac{4}{7}N\mu.
\label{surface_integral_high_g}
\end{align}
If $\mu/mg >H$, the upper integration limit is $z'=H$, giving
\begin{align}
    \int_S P(\mathbf{r}) \mathbf{r}\cdot\hat{\mathbf{n}}dS &= 2\int_0^{H}P(z)dV + P(H)V \nonumber \\
    &=2A\int_0^{H}P(z)dz +P(H)AH \nonumber \\
    &=\frac{Nmg/A}{1-(1-mgH/\mu)^{5/2}}\left[ 2A\int_0^H \left(1-\frac{mgz}{\mu}\right)^{5/2} dz +\left(1-\frac{mgH}{\mu}\right)^{5/2}AH\right] \nonumber \\
    &=\frac{4}{7}N\mu \left[\frac{1-(1-mgH/\mu)^{7/2}}{1-(1-mgH/\mu)^{5/2}}\right] 
     +NmgH\left[\frac{(1-mgH/\mu)^{5/2}}{1-(1-mgH/\mu)^{5/2}}\right].
\label{surface_integral_low_g}
\end{align}
The surface integrals in Eqs. (\ref{surface_integral_high_g}) and (\ref{surface_integral_low_g}) are equal to $2K-U$, where $K$ is given by Eq. (\ref{K_gravity}) and $U$ by Eq. (\ref{U_gravity}). This concludes the verification based on the virial theorem as given in Eq. (\ref{virial}).

\section{Conclusion}

We investigate the behavior of an ideal Fermi gas confined in a container subject to a uniform gravitational field at zero temperature. The semiclassical approximation is central to this work, enabling the derivation of some useful results. In particular, the semiclassical approach provides simple analytic expressions for the density profile and the chemical potential in a uniform gravitational field. It allows the identification of distinct weak- and strong-gravity regimes separated by a critical gravitational strength. Although the semiclassical description breaks down near the classical turning point, comparison with the exact quantum treatment demonstrates that it nevertheless yields reliable predictions for global quantities, such as the kinetic and potential energies, which are shown to be consistent with the virial theorem. The analysis presented in this article is intended for advanced theoretical studies of quantum many-body systems, particularly in contexts where semiclassical methods provide insight into underlying quantum behavior.

% \section*{Acknowledgments}
% Wittaya Kanchanapusakit acknowledges funding by Faculty of Science, King Mongkut's University of Technology Thonburi.

\section*{Author declarations}

The authors have no conflicts to disclose.


\begin{thebibliography}{99}
\bibitem{footnote1}
Note that this situation considered in this work differs from that of a typical metallic Fermi gas, since, in that case, electrostatic forces between the Fermi gas of electrons and the positive ions fix the volume that the gas will occupy. Instead, we consider a model for a noninteracting gas of fermions, which could approximate the behavior of dense cold neutrons.
\end{thebibliography}
\end{document}